\documentclass[prd,twocolumn,aps,showpacs,nofootinbib,nobibnotes,superscriptaddress,preprintnumbers,10pt]{revtex4-1}

\usepackage{amsmath}
\usepackage{graphicx}
\usepackage[colorlinks=true, allcolors=blue]{hyperref}

\newcommand{\D}{\mathrm{d}}
\newcommand{\E}{\mathrm{e}}
\newcommand{\I}{\mathrm{i}}

\begin{document}

\title{Kinetic Field Theory applied to Vector-Tensor Gravity}

\author{Lavinia Heisenberg} \email{lavinia.heisenberg@phys.ethz.ch}
\affiliation{Institute for Theoretical Physics, 
ETH Zurich, Wolfgang-Pauli-Strasse 27, 8093, Zurich, Switzerland}
 
\author{Matthias Bartelmann} \email{bartelmann@uni-heidelberg.de}
\affiliation{Universit\"at Heidelberg, Zentrum f\"ur Astronomie, Institut f\"ur Theoretische Astrophysik, Philosophenweg 12, 69120 Heidelberg, Germany}

\date{\today}

\begin{abstract}
The formation of cosmic structures is an important diagnostic for both the dynamics of the cosmological model and the underlying theory of gravity. At the linear level of these structures, certain degeneracies remain between different cosmological models and alternative gravity theories. It is thus indispensable to study the non-linear, late-time evolution of cosmic structures to try and disentangle their fundamental properties caused by the cosmological model or gravity theory itself. Conventionally, non-linear cosmic structure formation is studied by means of computationally expensive numerical simulations. Since these inevitably suffer from shot noise and are too time consuming to systematically scrutinize large parameter spaces of cosmological models or fundamental theories, analytical methods are needed to overcome the limitations of numerical simulations.

Recently, a new analytic approach to non-linear cosmic structure formation has been proposed based on kinetic field theory for classical particle ensembles. Within this theory, a closed, analytic, non-perturbative and parameter-free equation could be derived for the non-linear power spectrum of cosmic density perturbations which agrees very well with numerically simulated results to wave numbers $k\lesssim10\,h\,\mathrm{Mpc}^{-1}$ at redshift $z = 0$. In this Letter, we study for the first time the implications of alternative gravity theories for non-linear cosmic structure formation applying this promising new analytic framework. As an illustrative example, we consider vector-tensor theories, which support very interesting isotropic cosmological solutions.
\end{abstract}


\maketitle

\section{Introduction}

Cosmic structure formation is an indispensable process for probing into the true nature of gravity. The temperature and polarisation fluctuations in the cosmic microwave background (CMB) \cite{2016A&A...594A..13P, 2016A&A...594A...1P} as the earliest imprint of cosmic structures already place stringent constraints on alternative theories of gravity. The CMB temperature fluctuations were predominantly adiabatic and reflect fluctuations in the matter density. The dynamics of cosmic structures can be treated linearly as long as their relative density fluctuations remain smaller than the mean cosmic matter density. Linear analyses typically assume that the matter density and velocity fields are smooth and differentiable and governed by the hydrodynamical continuity and Euler equations \cite{2002PhR...367....1B}. Fluctuations in the gravitational potential are related to density fluctuations by the Poisson equation. For non-relativistic structures with length scales small compared to the Hubble radius, this simplified analytical description is appropriate.

Wide-area surveys reveal cosmic structures on all scales, marked by stars, galaxies, galaxy clusters, and extended filaments surrounding enormous voids. The cosmological standard model asserts that these structures originated from primordial density fluctuations. On small scales and at late cosmic times, their approximate description by linearized equations breaks down because the density fluctuations become highly non-linear, and the hydrodynamical equations fail because they cannot account for multi-valued velocity fields. This leads to the notorious shell-crossing problem. Conventionally, therefore, non-linear cosmic structure formation is studied using highly elaborate $N$-body simulations. They decompose the cosmic density field into pseudo-particles with a mass set by the spatial resolution of the simulation. Starting from given initial conditions, the gravitational interaction of the matter particles is simulated with various approximation schemes. Simulations with sufficiently high resolution reveal gravitationally bound structures with universal properties.

Testing alternative theories of gravity on cosmological scales requires a thorough comparison of late-time and non-linear cosmic structures to the standard evolution in the framework of general relativity \cite{2018LRR....21....2A, 2013LRR....16....6A}. Therefore, it is an indispensable task to achieve a fundamental understanding of non-linear cosmic structure formation in order to sensitively test the cosmological consequences of gravity theories. In most alternative theories of gravity, the time sequence of gravitational clustering, the evolution of peculiar velocities and the number density of collapsed objects can be expected to be significantly influenced. Alterations in the gauge-invariant matter-density contrast and its relation to the gravitational potential, in the gravitational slip parameter, the effective gravitational potential, and in the growth rate of cosmic structures can then potentially be discovered in the data provided by large-scale structure observations. However, high-resolution simulations of cosmic structures in the sufficiently deeply non-linear regimes for a large variety of alternative theories of gravity would be forbiddingly time consuming and unaffordable. Hence, it is crucial to develop analytical approaches to non-linear cosmic structure formation. See \cite{2018arXiv180701725H} for a recent review on the systematic construction of alternative gravity theories.

In this Letter, we will for the first time generalize and apply recent results obtained within kinetic field theory (KFT) \cite{2016NJPh...18d3020B, 2017NJPh...19h3001B, 2018JSMTE..04.3214F, 2018JPhCo...2b5020D, 2015PhRvE..91f2120V} (see the review article \cite{2019arXiv190501179B}) to alternative theories of gravity. As a representative class of such alternatives, we choose generalised Proca theories \cite{2014JCAP...05..015H, 2016PhLB..757..405B, 2014JHEP...04..067T, 2016JCAP...02..004A}, but our analysis can be easily extended to other classes of alternative theories. Comparable to numerical simulations, KFT decomposes the cosmic matter field into classical pseudo-particles whose trajectories in phase space are governed by the Hamiltonian equations of motion. The Hamiltonian equations admit the construction of a (retarded) Green's function \cite{2015PhRvD..91h3524B}. Note that the notorious shell- or stream-crossing problem does not even occur in this approach since Hamiltonian trajectories cannot cross in phase space.

\section{Kinetic field theory in essence}

As mentioned above, KFT dissolves the cosmic matter field into an ensemble of classical particles in phase space. Their initial state is statistically characterized by a distribution of appropriately correlated phase-space positions at a sufficiently early time, for example at the time of matter-radiation decoupling. At this time, the particle ensemble can be drawn from a homogeneous and isotropic Gaussian random field \cite{2016A&A...594A..16P, 2016A&A...594A..17P}. The Hamiltonian phase-space trajectories of the particles are described by a retarded Green's function $G(t,t')$.

As in any non-equilibrium, statistical field theory, the central mathematical object of KFT is a generating functional
\begin{equation}
  Z = \int\mathcal{D}[\varphi]\int\mathcal{D}[\varphi_i]\,
  P(\varphi|\varphi_i)P(\varphi_i)
\end{equation} 
for field configurations $\varphi$. Their probability distribution is specified by the distribution $P(\varphi_i)$ for the initial configuration $\varphi_i$ at time $t_i=0$, and the transition probability $P(\varphi|\varphi_i)$ from $\varphi_i$ to a configuration $\varphi$ at any later time $t>t_i$. For classical point particles, the path integrations simply turn into ordinary integrations over phase-space coordinates $x = (q,p)$. Since classical particle trajectories are deterministic, the transition probability is given by a functional delta distribution. With $G(t,t')$ known, the particle trajectories beginning at the initial phase-space position $x_i$ are formally given by
\begin{equation}
  \bar x(t) = G(t,0)x_i-\int_0^td t'\,G(t,t')\nabla V(t')\;,
\end{equation}
where $V$ is the interaction potential between the particles. Further introducing a generator field $J$ for later functional derivatives, the generating functional then turns into
\begin{align}
  Z[J] &= \int\D x_i\int\D x\,\delta\left(x-\bar x\right)\,P(x_i)\,
  \E^{\I\int_0^t\D t'\langle J,x\rangle} \nonumber\\ &=
  \int\D\Gamma\,\E^{\I\int_0^t\D t'\langle J,\bar x\rangle}\;,
\end{align} 
where $\langle\cdot,\cdot\rangle$ is a suitably defined scalar product between the generator field $J$ and the coordinates $\bar x$ of all particles in the ensemble. We abbreviate the initial phase-space measure by $\D\Gamma := P(x_i)\D x_i$.

For an ensemble of $N$ point particles, the matter density in configuration space is determined by a sum over delta distributions. In a Fourier representation,
\begin{equation}
  \rho(1) := \rho\left(\vec k_1,t_1\right) =
  \sum_{i=1}^N\E^{-\I\vec k_1\cdot\vec q_i(t_1)}\;.
\end{equation}
The position $\vec q_i(t_1)$ of the $i$-th particle at time $t_1$ can be obtained from the generating functional by a functional derivative with respect to the component $J_{q_i}(t_1)$ of the generator field. Thus, the average matter density is obtained from the generating functional by applying the operator
\begin{equation}
  \hat\rho(1) = \sum_{i=1}^N\hat\rho_i(1) =
  \sum_{i=1}^N
  \exp\left(-\I\vec k_1\cdot\frac{\delta}{\I\delta J_{q_i}(t_1)}\right)
\end{equation}
to $Z[J]$. Having taken all functional derivatives required, the generator field is set to zero.

Since derivatives generate infinitesimal translations, exponentials of derivatives generate finite translations. Thus, the density operator $\hat\rho(1)$ creates a shift
\begin{equation}
  L(1) = -\vec k_1\,\delta(t-t_1)\left(\begin{matrix}0\\1\end{matrix}\right)
\end{equation} 
and has the effect
\begin{equation}
  \left.\hat\rho(1)Z[J]\right|_{J=0} = \langle\rho(1)\rangle =
  \int\D\Gamma\,\E^{\I\langle L(1),\bar x\rangle}
\end{equation}
on the generating functional. Applying $n$ density operators to $Z$ returns the $n$-point correlation function
\begin{equation}
  G_n = \langle\rho(1)\ldots\rho(n)\rangle =
  \left.\prod_{i=1}^n\hat\rho(i)Z[J]\right|_{J = 0}\;.
\end{equation}

The initial phase-space distribution of the particle ensemble can be drawn from a Gaussian random field, which is completely specified by the density-fluctuation power spectrum. It is convenient to introduce comoving spatial coordinates and to replace the cosmological time by the linear growth factor $D_+$ of the density fluctuations, $t\to\tau := D_+(t)-D_+(t_i)$.

The interaction between any two particles at positions $1$ and $2$ is described by the potential
\begin{equation}
  v(12) = \int_k\,\tilde v(k)\,
  \E^{\I\vec k\cdot(\vec q_1(\tau_1)-\vec q_2(\tau_1))}\;.
\end{equation} 
Since we can extract the particle positions from the generating functional by functional derivatives with respect to the generator-field components $J_{q_1}$ and $J_{q_2}$, the interaction potential can be turned into an interaction operator appearing in an exponential function. The Taylor expansion of this exponential operator then corresponds to the conventional perturbative approach in terms of Feynman diagrams \cite{2016NJPh...18d3020B, 2018JSMTE..04.3214F}.

In a different, non-perturbative approximation, the interaction term $S_I$ between all particle pairs is replaced by its average $\langle S_I\rangle$ weighted with the two-point correlation function $\xi(12) := \xi(|\vec q_1-\vec q_2|)$ of the particles. This mean-field approach allows the derivation of the expression
\begin{equation}
  \mathcal{P}(k,\tau) = \E^{-Q_0+\I\langle S_I\rangle}
  \int\D^3q\,\left(\E^{Q(q)}-1\right)\E^{\I\vec k\cdot\vec q}
\label{eq:nlP}
\end{equation} 
for the non-linear density-fluctuation power spectrum \cite{2017arXiv171007522B}. Here,
\begin{equation}\label{expressionQq}
  Q(q) := -g_{qp}^2(\tau,0)\,k^2\,a_\parallel(q)
\end{equation} 
with the correlation function $a_\parallel(q)$ of momentum components parallel to the wave vector $\vec k$ between particles separated by $q$. Furthermore, $Q_0 = Q(0)$, and $g_{qp}(\tau,\tau')$ is the position-momentum component of the retarded Green's function.

Within the realm of general relativity, this mean-field approach was applied to a standard, $\Lambda$CDM model universe \cite{2017arXiv171007522B}. The resulting density-fluctuation power spectrum according to (\ref{eq:nlP}) was shown to agree very well with the result from numerical simulations to $k\lesssim10\,h\,\mathrm{Mpc}^{-1}$ at redshift $z=0$. In this Letter, we show that the KFT approach is not just limited to the standard gravity theory and the standard model of cosmology, but can easily be generalized to a variety of different field theories for gravity. In a first illustrative application, we focus on a general class of vector-tensor theories in the following section.

\section{Generalized Proca theories}

In this Letter, we apply the KFT approach to generalised Proca theories of gravity. These are the most general vector-tensor theories with second order equations of motion, where the vector field possesses only three propagating degrees of freedom. The action of these generalized Proca theories is given by \cite{2014JCAP...05..015H, 2016PhLB..757..405B}
\begin{equation}
  S = \int\D^4x \sqrt{-g} \left( {\cal L}
  +{\cal L}_M \right)\;,\quad
  {\cal L}=\sum_{i=2}^{6} {\cal L}_i\;,
\label{LagProca}
\end{equation}
where $g$ is the determinant of the metric tensor $g_{\mu \nu}$ and ${\cal L}_M$ represents the matter Lagrangian density. Following the standard definition, we denote the field strength by $F_{\mu \nu}=\nabla_{\mu}A_{\nu}-\nabla_{\nu}A_{\mu}$ and its dual by $\tilde{F}^{\mu \nu}=\epsilon^{\mu \nu \alpha \beta}F_{\alpha \beta}/2$, and we define the operator $K_{\mu\nu}=\nabla_{\mu}A_{\nu}$ for convenience. The Lagrangian densities of the generalized Proca action are given by \cite{2014JCAP...05..015H, 2016PhLB..757..405B}
\begin{align}
\label{gen_ProcaLag}
  {\cal L}_2 &= G_2(X,F,Y)\;, \\
  {\cal L}_3 &= G_3(X) [K]\;, \\
  {\cal L}_4 &= G_4(X)R+G_{4,X} \left[ [K]^2 -[K^2] \right]\;, \\
  {\cal L}_5 &= G_{5}(X) G_{\mu \nu}K^{\mu\nu}-
    \frac{G_{5,X}}{6}\left[[K]^2-3[K][K^2]+2[K^3]\right] \nonumber \\
    &-g_5(X) \tilde{F}^{\alpha\mu}\tilde{F}^\beta_{\phantom{\beta}\mu}
    K_{\alpha\beta}\;, \\
  {\cal L}_6 &= G_6(X) L^{\mu \nu \alpha \beta}
    K_{\mu\nu} K_{\alpha\beta}
    +\frac{G_{6,X}}{2}\tilde{F}^{\alpha \beta} \tilde{F}^{\mu \nu}
    K_{\alpha\mu} K_{\beta\nu}\;.
\end{align}
The Lagrangian ${\cal L}_2$ can be an arbitrary function of the three independent contractions  
$X =-\frac12 A_{\mu} A^{\mu}$, 
$F = -\frac14 F_{\mu \nu} F^{\mu \nu}$ and 
$Y = A^{\mu}A^{\nu} {F_{\mu}}^{\alpha}F_{\nu \alpha}$,
whereas the functions $G_{3,4,5,6}$ and $g_5$ can only depend on $X$. Note that $G_{i,X} \equiv \partial G_{i}/\partial X$ denotes the partial derivatives of the corresponding functions. As one can see, the vector field directly couples non-minimally to the Lovelock invariants and their corresponding equivalents at the level of the equations of motion in terms of divergence-less tensors. One unique and important coupling happens via the double dual Riemann tensor 
$L^{\mu \nu \alpha \beta}=\frac14 \epsilon^{\mu \nu \rho \sigma} \epsilon^{\alpha \beta \gamma \delta} R_{\rho \sigma \gamma \delta}$,
with the Levi-Civita tensor $\epsilon^{\mu \nu \rho \sigma}$ and the Riemann tensor $R_{\rho \sigma \gamma \delta}$. Note that the non-minimal couplings introduced here are crucial for the nature of second-order equations of motion. The interactions in ${\cal L}_6$ and the interaction proportional to $g_5(X)$ are purely intrinsic vector interactions with no corresponding scalar counterpart. 

We begin with (\ref{LagProca}) and specialize the theory to an unperturbed Friedmann background. The line element is then $\D s^2 = -N^2(t)\D t^2+a^2(t)\D\vec x^{\,2}$ with the lapse function $N$ and the scale factor $a$. Compatible with the background symmetry, the vector field $A$ can acquire the only allowed field configuration $A^\mu=(A^0(t),0,0,0)$, where we chose the time component as $A^0 = \phi(t)/N(t)$. The corresponding Hamiltonian of the unperturbed background $ \mathcal{H} =\Pi^\mu \dot{\mathcal{O}}_\mu- \mathcal{L}$, with $\mathcal{O}=(N,a,\phi)$, and the conjugate momenta $\Pi^\mu=\frac{\partial \mathcal{L}}{\partial \dot{\mathcal{O}}_\mu}$, are simply given by
\begin{equation}
 \mathcal{H} = -Na^3(G_2+6H^2G_4-6G_{4,X}H^2\phi^2+2G_{5,X}H^3\phi^3)\,.
\end{equation}
The Einstein field equations for the background can be expressed as
\begin{align}
  3M_{\rm pl}^2H^2 &= \rho_{\rm DE}+\rho_M \nonumber\\
  M_{\rm pl}^2 (3H^2+2\dot{H}) &= -P_{\rm DE}-P_M
\end{align}
with the Hubble function $H=\dot{a}/a$. We have further assumed $G_4=M_{\rm pl}^2/2+g_4$ and introduced the effective energy density and the pressure of the vector field \cite{2016JCAP...06..048D}
\begin{align}
  \rho_{\rm DE} &=-G_2+G_{2,X}\phi^2+3G_{3,X}\phi^3+
  12G_{4,X}H^2\phi^2-6g_4H^2
  \nonumber\\ &+6G_{4,XX}\phi^2H^2\phi^2-
  G_{5,XX}H^3\phi^5-5G_{5,X}H^3\phi^3\;,\nonumber\\
  P_{\rm DE} &= G_2-G_{3,X}\phi^2\dot\phi-
  2G_{4,X}\phi\left(3H^2\phi+2H\dot\phi+2\dot H\phi\right)
  \nonumber\\ &-
  4G_{4,XX}H\phi^3\dot\phi+G_{5,XX}H^2\phi^4\dot\phi+2g_4(3H^2+\dot{H})\nonumber\\
  &+G_{5,X}H\phi^2\left(2\dot H\phi+2H^2\phi+3H\dot\phi\right)\;.
\label{eq:9}
\end{align}
Similarly, the scalar field equation reads
\begin{align}
  0 &= \phi\,\Bigl\{
    G_{2,X}+3HG_{3,X}\phi+6H^2\left(G_{4,X}+G_{4,XX}\phi^2\right)
    \nonumber\\ &-
    H^3\left(3G_{5,X}+G_{5,XX}\phi^2\right)\phi
  \Bigr\} = 0\;.
\label{eq:7}
\end{align}
Well equipped with these equations, we can proceed to solve for the background evolution. The relevant dynamical field is the modified Hubble function $H(t)$. The temporal component of the vector field is just an auxiliary field and can be integrated out using equation \eqref{eq:7}.

We shall consider small perturbations on top of the homogeneous and isotropic background, $g_{\mu\nu}=\bar{g}_{\mu\nu}+\delta g_{\mu\nu}$ and $A_\mu=\bar{A}_\mu+\delta A_\mu$, and express the perturbations in terms of the irreducible representations of the underlying background symmetries. The key equations for us are the modified Poisson equation and the growth equation. The gauge-invariant Bardeen potential $\Psi$ satisfies the Poisson equation $-k^2\Psi = 4\pi G_\mathrm{eff}a^2\rho\delta$, sourced by the gauge-invariant matter-density contrast $\delta=\frac{\delta\rho}{\rho}$. For non-relativistic matter perturbations much smaller than the sound horizon, the density contrast satisfies the well-known perturbation equation $\ddot\delta+2H\dot\delta-4\pi G_\mathrm{eff}\rho\delta = 0$, where the effective gravitational constant
\begin{equation}
  G_\mathrm{eff} = \frac{H}{4\pi\phi}\frac{\mu_2\mu_3-\mu_1\mu_4}{\mu_5}
\label{eq:13}
\end{equation}
appears. For brevity, we omit the exact expressions for the functions $\mu_i$ here. They can be found in Eqs.\ (5.10)-(5.14) and (5.18) in \cite{2016PhRvD..94d4024D}.

\section{KFT applied to generalized Proca theories}

Having established the fundamental equations above, adapting KFT with the mean-field approximation to generalized Proca theories is quite straightforward. We solely need to specify
\begin{itemize}
  \item the background evolution in generalized Proca theories, characterised by the Hubble function $H$;
  \item the time evolution of the effective gravitational constant $G_\mathrm{eff}$;
  \item the growth factor $D_+$ for linear perturbations; and
  \item possible changes to the gravitational potential.
\end{itemize}
The first point requires a concrete dark energy model. A specific and promising, yet simple model was described in \cite{2016PhRvD..94d4024D}. It sets the functions $G_i$ with $i = 2\ldots5$ to
\begin{align}
  G_2 &= b_2X^{p_2}+F\;, \nonumber\\
  G_3 &= b_3X^{p_3}\;, \nonumber\\
  G_4 &= \frac{M_\mathrm{Pl}^2}{2}+b_4X^{p_4}\;,\quad\mbox{and} \nonumber\\
  G_5 &= b_5X^{p_5}\;,
\end{align}
and allows solutions of the form $\phi^p \propto H^{-1}$. The background equations can be easily solved and effectively recast into the form
 \begin{equation}
  \frac{\Omega_\mathrm{DE}}{\Omega_\mathrm{DE,0}} =
  \frac{\Omega_\mathrm{r}}{\Omega_\mathrm{r,0}}\,a^{4(1+s)}\;,
\label{eq:22}
\end{equation}
with $s := p_2/p$. For this specific dark-energy model, we can write the growth equation as
\begin{equation}\label{eomdelta}
  \delta''+\frac{1+(3+4s)\Omega_\mathrm{DE}}{2(1+s)\Omega_\mathrm{DE}}\delta'-
  \frac{3}{2}\frac{G_\mathrm{eff}}{G}(1-\Omega_\mathrm{DE})\delta=0\;.
\end{equation}

\begin{figure}[!ht]
  \includegraphics[width=\hsize]{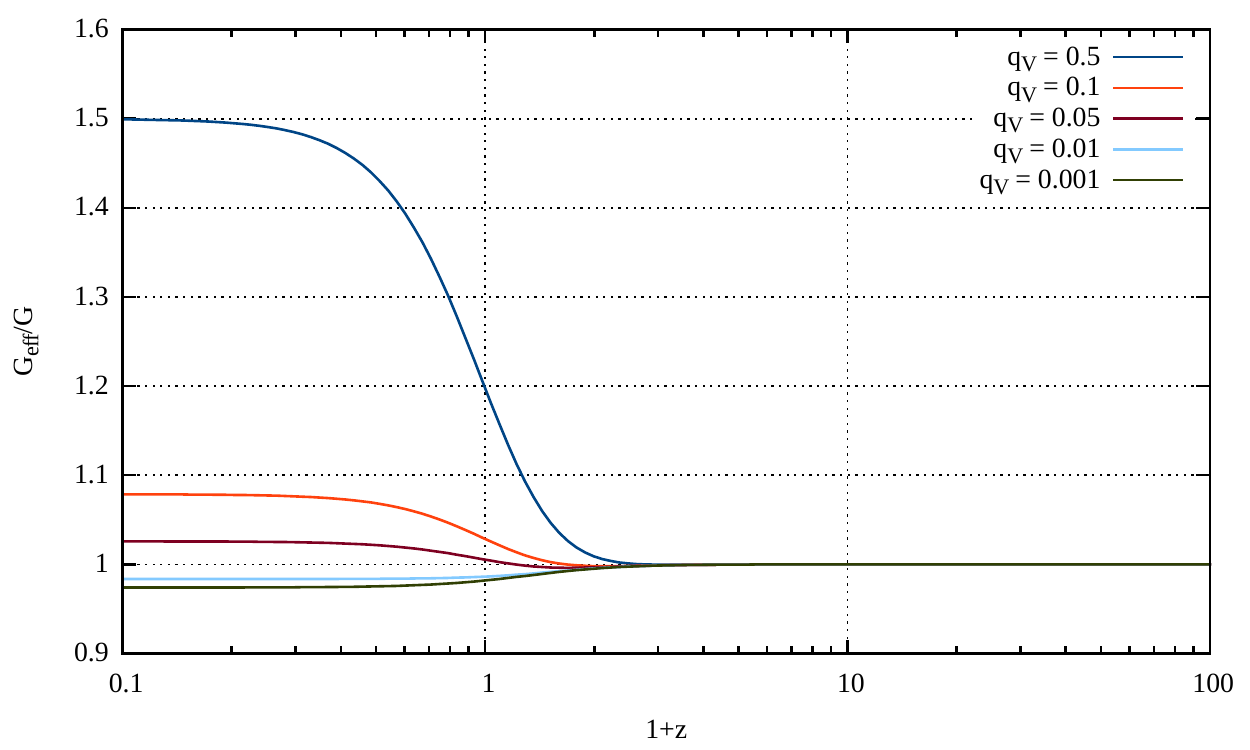}
\caption{The evolution of the effective gravitational coupling $G_\mathrm{eff}$ as a function of redshift $z$ for different values of $q_v$. The model parameters have been chosen as in \cite{2016PhRvD..94d4024D}, where we refer the reader for more detail. The parameters quantifying the non-minimal couplings are $\beta_4=10^{-4}$ and $\beta_5=0.052$.}
\label{fig_geff}
\end{figure}

The evolution of the effective gravitational coupling $G_\mathrm{eff}$ in Eq.\ \eqref{eq:13} is shown in Fig.\ \ref{fig_geff} for the model parameters represented by $\beta_i := \frac{b_ip_i}{2^{p_i-p_2}b_2p_2}$ used in \cite{2016PhRvD..94d4024D}. It is shown there for different values for $q_v$, which stands for the redressing of the kinetic term of the vector perturbations,
\begin{equation}
  q_\mathrm{V} =
  G_{2,F}+2G_{2,Y}\phi^2-4g_5H\phi+2G_6H^2+2G_{6,X}H^2\phi^2\;,
\end{equation} 
and enters in a very specific way into the functions $\mu_i$ in Eq.\ \eqref{eq:13} (see \cite{2016PhRvD..94d4024D}). This parameter encodes the effects of the intrinsic vector modes. The effective gravitational coupling obtained agrees with Fig.\ 1 of \cite{2016PhRvD..94d4024D} as we have chosen the same parameter set. Note that this model contains non-minimal couplings with non-vanishing $\beta_4$ and $\beta_5$ coming from the presence of $G_4$ and $G_5$. 

The vector-tensor modification of gravity specified in this and the preceding sections enters into KFT in the following way:
In \eqref{eq:nlP}, the mean interaction term $\langle S_I\rangle$ contains the gravitational potential between particles, whose amplitude is modified by the effective gravitational constant $G_\mathrm{eff}$ given in \eqref{eq:13}. The time coordinate $\tau$ is conveniently chosen to be the growth factor of linear density perturbations, which is the growing solution of \eqref{eomdelta}. Similarly, the time evolution of the background, expressed by the Hubble function $H$, affects the propagators $g_{qp}$ appearing in the function $Q(q)$ and its specialization $Q_0$ in \eqref{eq:nlP}. The power spectrum of the initial, linear density and momentum perturbations is unchanged compared to general relativity and is set to the standard, $\Lambda$CDM power spectrum here. It enters into KFT via the momentum correlation function $a_\parallel(q)$ entering into \eqref{eq:nlP} via the function $Q(a)$ from \eqref{expressionQq}.

\begin{figure}[!ht]
  \includegraphics[width=\hsize]{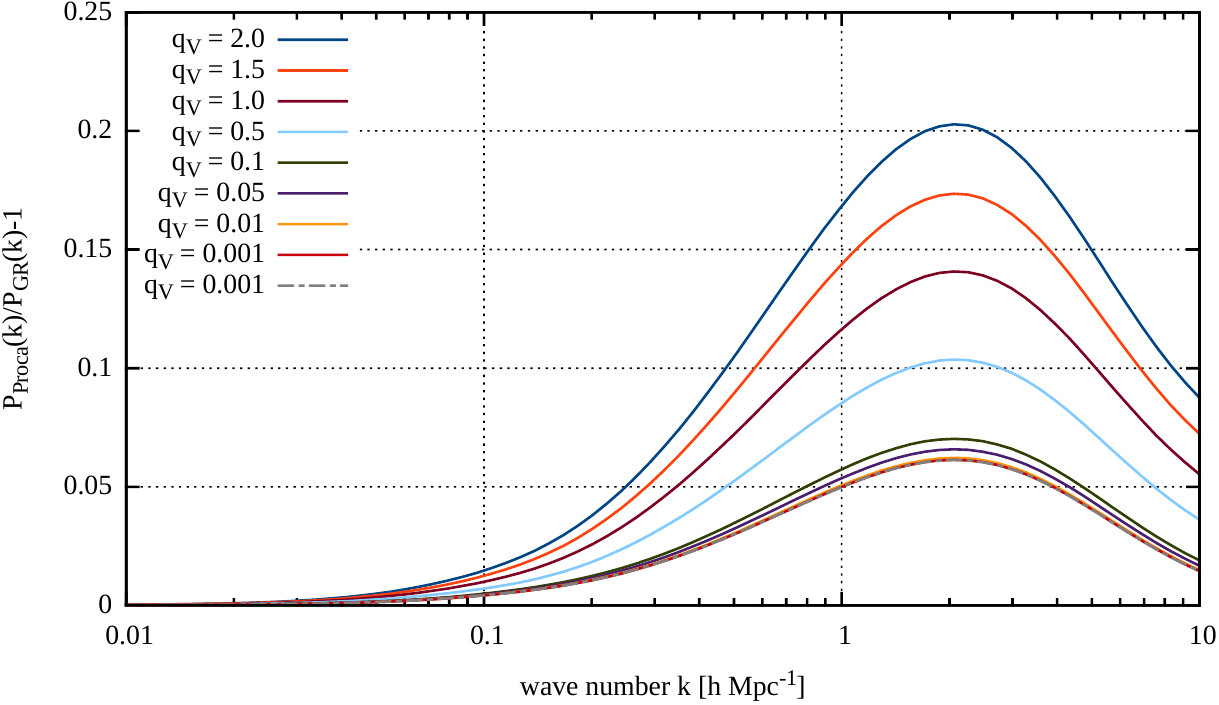}
\caption{Change of the non-linear matter power spectrum relative to its behaviour in general relativity and the standard cosmological model as a function of the wave number $k$ for a generalized Proca theory with the parameters as chosen for Fig.\ \ref{fig_geff} following \cite{2016PhRvD..94d4024D}.}
\label{fig_ps_m1}
\end{figure}

As mentioned above, the relative difference between the local density and the mean density is the density contrast. Its statistical properties as a function of scale are described by the matter power spectrum, which is the Fourier transform of the matter correlation function as a function of the wave number $k$. The quantity $\mathcal{P}(k,\tau)$ is the power spectrum according to KFT, non-linearly evolved within the mean-field approximation of the interaction term $\langle S_I\rangle$. Figure \ref{fig_ps_m1} shows the change of the non-linear power spectrum relative to general relativity and the standard cosmological model for the generalized Proca theory with the same parameter choice as for Fig.\ \ref{fig_geff}. The figure clearly shows an increase of power peaking at scales near $k\sim 2\,h\,\mathrm{Mpc}^{-1}$, reaching $\sim20\,\%$ for $q_v = 2$. Interestingly, this enhanced power corresponds to a higher number density of cosmic structures on the scale of $\sim 3\,h^{-1}\,\mathrm{Mpc}$, characteristic of massive galaxy clusters.

\begin{figure}[!ht]
  \includegraphics[width=\hsize]{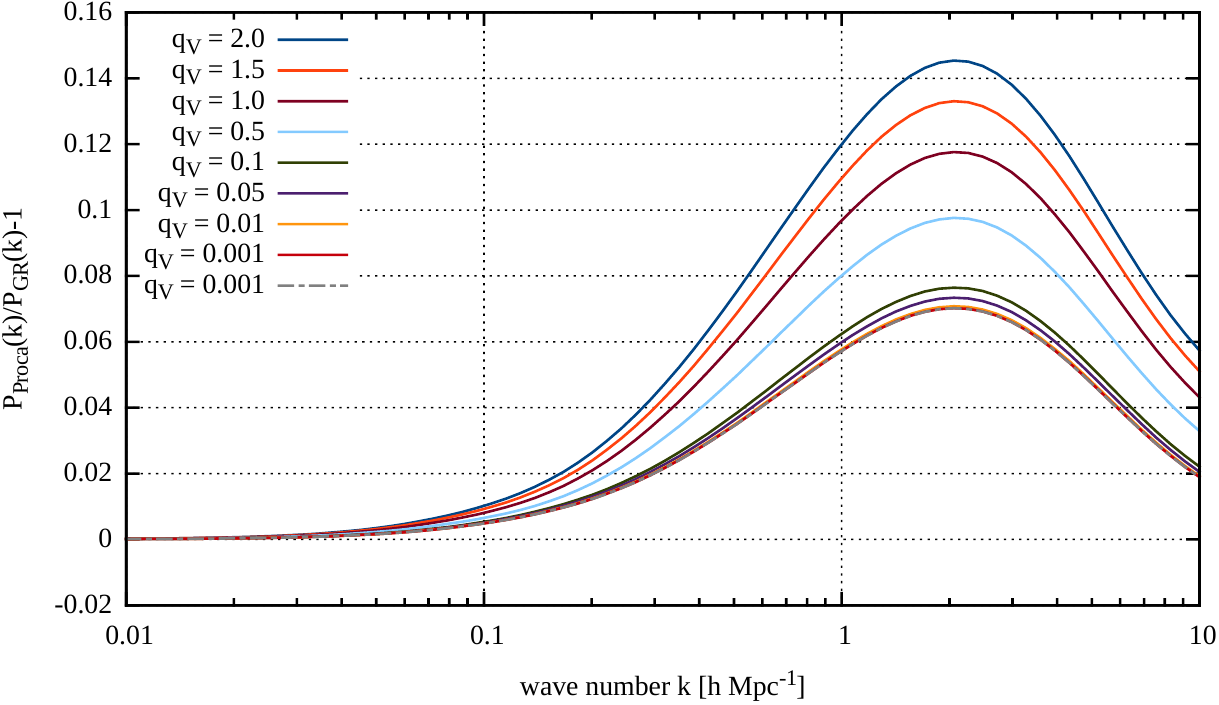}
\caption{As in Fig.\ \ref{fig_ps_m1}, this Figure shows the change of the non-linear matter power spectrum relative to the standard cosmological model in general relativity for a generalized Proca theory. Compared to Fig.\ \ref{fig_ps_m1}, the parameters are now constrained by the empirical limit $|c_T/c-1|<10^{-15}$ on the propagation speed of gravitational waves, reflected by $\beta_4 = 0 = \beta_5$. All other parameters remain unchanged from their values chosed for Fig.\ \ref{fig_geff} following \cite{2016PhRvD..94d4024D}.}
\label{fig_ps_m2}
\end{figure}

The breathtaking discovery of two merging neutron stars has strictly constrained the propagation speed $c_T$ of gravitational waves relative to the speed of light by $|\frac{c_T}{c}-1|<10^{-15}$. Thus, any modification of gravity implying a propagation speed of gravitational waves different from the speed of light is tightly constrained. Among the interactions of the generalized Proca theory (\ref{gen_ProcaLag}), the non-minimal couplings to the gravity sector in Lagrangians $\mathcal{L}_4$ and $\mathcal{L}_5$ contribute to the tensor propagation speed as follows \cite{2016JCAP...06..048D, 2016PhRvD..94d4024D}
\begin{equation}
  c_T^2 = \frac{2G_4+\phi^2\dot{\phi}G_{5,X}}
    {2G_4-2\phi^2G_{4,X}+H\phi^3G_{5,X}} \;.
\end{equation}
Evidently, this expression demands $G_{4,X}=0$ and $G_5=0$ for $c_T=1$. To illustrate the effect of this constraint, we now consider a second generalized Proca model with $\beta_4=0$ and $\beta_5=0$, and with the remaining parameters untouched. The corresponding non-linear power spectrum is shown in Fig.\ \ref{fig_ps_m2}.

The power is still enhanced with a peak at scales near $k\sim 2\,h\,\mathrm{Mpc}^{-1}$, but with a slightly reduced amplitude. This implies that in dark-energy models based on generalized Proca theories with parameters compatible with the measured propagation speed of gravitational waves, the increase of structure formation near the scale of massive galaxy clusters is not quite as strong as in models with $\beta_4\ne0$ and $\beta_5\ne0$. Even though the latter models allow a decrease of structure formation at the linear order for small $q_v$ (see Fig.\ \ref{fig_geff} with $q_v=0.01$ and $q_v=0.001$), this is completely lost when non-linear evolution is taken into account. Interestingly, numerical $N$-body simulations find very similar results with a peak at slightly larger scales in other examples of modified-gravity theories \cite{2014MNRAS.440...75B, 2018A&A...619A..38P}

For both models studied here, the amplitude of observable structures near the scale of massive galaxy clusters would be enhanced, which could possibly be directly tested using non-linear probes of structure formation, such as galaxy clusters selected for their strong gravitational lensing or thermal Sunyaev-Zel'dovich effects or their hot and bright X-ray emission.

Note that the power spectra shown relative to their standard counterpart were calculated within KFT with the mean-field approximation for the interaction term. For proceeding towards more deeply non-linear scales, where the Vainshtein screening mechanism will become relevant \cite{2016PhRvD..93j4016D}, we would need to go beyond this approximation and take into account the non-linear interactions of the vector field. This is clearly beyond the scope of the present work and will be intensively studied in future work.

\section{Conclusion}

In this Letter, we have for the first time applied the newly developed kinetic field theory (KFT) for non-linear cosmic structure formation to alternative gravity theories. As a representative theory, we have chosen the generalized Proca theories. They contain an additional vector field in the gravity sector with non-minimal couplings to the tensor field. We have first investigated the modified background evolution in these theories and derived the corresponding Hubble function $H(t)$ and the Hamiltonian $ \mathcal{H}(t)$ of the background fields. We then deduced from the linear perturbations the form of the effective gravitational coupling constant $G_\mathrm{eff}$ together with the growth factor $D_+$ for linear perturbations and the changes to the gravitational potential. In a second step, we have directly applied the KFT approach in the mean-field approximation of the interaction term, and obtained the non-linear cosmic density-fluctuation power spectrum in this approximation. We have chosen two different parameter sets for the dark-energy models within the class of generalized Proca theories. The first contains explicit non-minimal couplings to gravity, while the second omits these couplings in order to be compatible with the recent constraint on the speed of gravitational waves. In both cases, non-linear power is increased with a peak near scales $k\sim 2\,h\,\mathrm{Mpc}^{-1}$. It would be very interesting to implement such models in an $N$-body simulation \cite[see][for examples]{2014MNRAS.440...75B, 2018A&A...619A..38P} and to compare the power spectra found there with those obtained here. For proceeding further into the non-linear regime of cosmic structure formation, it will also be crucial to consider the vector interactions beyond the mean-field approximation and investigate the effect of the Vainshtein mechanism.

\section*{Acknowledgements}

LH is supported by funding from the European Research Council (ERC) under the
European Union’s Horizon 2020 research and innovation programme
grant agreement No 801781 and by the Swiss National
Science Foundation grant 179740.

\bibliographystyle{apsrev4-1}
\bibliography{KFT_VT}

\end{document}